\documentstyle[aps,preprint,graphicx]{revtex}
\tighten
\draft

\begin{document}

%%%%%%%%%%%%%%%%%%%%%%%%%%%%%%%%%%%%%%%%%%%%%%%%%%%

\newcommand{\sst}[1]{{\scriptscriptstyle #1}}
\newcommand{\beq}{\begin{equation}}
\newcommand{\eeq}{\end{equation}}
\newcommand{\beqa}{\begin{eqnarray}}
\newcommand{\eeqa}{\end{eqnarray}}
\newcommand{\dida}[1]{/ \!\!\! #1}
\renewcommand{\Im}{\mbox{\sl{Im}}}
\renewcommand{\Re}{\mbox{\sl{Re}}}
\def\simge{\hspace*{0.2em}\raisebox{0.5ex}{$>$}
     \hspace{-0.8em}\raisebox{-0.3em}{$\sim$}\hspace*{0.2em}}
\def\simle{\hspace*{0.2em}\raisebox{0.5ex}{$<$}
     \hspace{-0.8em}\raisebox{-0.3em}{$\sim$}\hspace*{0.2em}}
\def\dn{{d_n}}
\def\de{{d_e}}
\def\datom{{d_{\sst{A}}}}
\def\grhobar{{{\bar g}_\rho}}
\def\gpibar{{{\bar g}_\pi^{(I) \prime}}}
\def\gpibarz{{{\bar g}_\pi^{(0) \prime}}}
\def\gpibaro{{{\bar g}_\pi^{(1) \prime}}}
\def\gpibart{{{\bar g}_\pi^{(2) \prime}}}
\def\mn{{m_{\sst{N}}}}
\def\mx{{M_X}}
\def\mrho{{m_\rho}}
\def\qpv{{Q_{\sst{W}}}}
\def\lamtv{{\Lambda_{\sst{TVPC}}}}
\def\lamtvs{{\Lambda_{\sst{TVPC}}^2}}
\def\lamtvc{{\Lambda_{\sst{TVPC}}^3}}

%     \hspace{-0.8em}\raisebox{-0.3em}{$\sim$}\hspace*{0.2em}}
\def\bra#1{{\langle#1\vert}}
\def\ket#1{{\vert#1\rangle}}
\def\coeff#1#2{{\scriptstyle{#1\over #2}}}
\def\undertext#1{{$\underline{\hbox{#1}}$}}
\def\hcal#1{{\hbox{\cal #1}}}
\def\sst#1{{\scriptscriptstyle #1}}
\def\eexp#1{{\hbox{e}^{#1}}}
\def\rbra#1{{\langle #1 \vert\!\vert}}
\def\rket#1{{\vert\!\vert #1\rangle}}

\def\lsim{{ <\atop\sim}}
\def\gsim{{ >\atop\sim}}
\def\nubar{{\bar\nu}}
\def\psibar{{\bar\psi}}
\def\Gmu{{G_\mu}}
\def\alr{{A_\sst{LR}}}
\def\wpv{{W^\sst{PV}}}
\def\evec{{\vec e}}
\def\notq{{\not\! q}}
\def\notl{{\not\! \ell}}
\def\notk{{\not\! k}}
\def\notp{{\not\! p}}
\def\notpp{{\not\! p'}}
\def\notder{{\not\! \partial}}
\def\notcder{{\not\!\! D}}
\def\notA{{\not\!\! A}}
\def\notv{{\not\!\! v}}
\def\Jem{{J_\mu^{em}}}
\def\Jana{{J_{\mu 5}^{anapole}}}
\def\nue{{\nu_e}}
\def\mn{{m_{\sst{N}}}}
\def\mns{{m^2_{\sst{N}}}}
\def\me{{m_e}}
\def\mes{{m^2_e}}
\def\mq{{m_q}}
\def\mqs{{m_q^2}}
\def\mw{{M_{\sst{W}}}}
\def\mz{{M_{\sst{Z}}}}
\def\mzs{{M^2_{\sst{Z}}}}
\def\ubar{{\bar u}}
\def\dbar{{\bar d}}
\def\sbar{{\bar s}}
\def\qbar{{\bar q}}
\def\sstw{{\sin^2\theta_{\sst{W}}}}
\def\gv{{g_{\sst{V}}}}
\def\ga{{g_{\sst{A}}}}
\def\pv{{\vec p}}
\def\pvs{{{\vec p}^{\>2}}}
\def\ppv{{{\vec p}^{\>\prime}}}
\def\ppvs{{{\vec p}^{\>\prime\>2}}}
\def\qv{{\vec q}}
\def\qvs{{{\vec q}^{\>2}}}
\def\xv{{\vec x}}
\def\xpv{{{\vec x}^{\>\prime}}}
\def\yv{{\vec y}}
\def\tauv{{\vec\tau}}
\def\sigv{{\vec\sigma}}

\def\sst#1{{\scriptscriptstyle #1}}
\def\gpnn{{g_{\sst{NN}\pi}}}
\def\grnn{{g_{\sst{NN}\rho}}}
\def\gnnm{{g_{\sst{NNM}}}}
\def\hnnm{{h_{\sst{NNM}}}}
\def\xivz{{\xi_\sst{V}^{(0)}}}
\def\xivt{{\xi_\sst{V}^{(3)}}}
\def\xive{{\xi_\sst{V}^{(8)}}}
\def\xiaz{{\xi_\sst{A}^{(0)}}}
\def\xiat{{\xi_\sst{A}^{(3)}}}
\def\xiae{{\xi_\sst{A}^{(8)}}}
\def\xivtez{{\xi_\sst{V}^{T=0}}}
\def\xivteo{{\xi_\sst{V}^{T=1}}}
\def\xiatez{{\xi_\sst{A}^{T=0}}}
\def\xiateo{{\xi_\sst{A}^{T=1}}}
\def\xiva{{\xi_\sst{V,A}}}
\def\rvz{{R_{\sst{V}}^{(0)}}}
\def\rvt{{R_{\sst{V}}^{(3)}}}
\def\rve{{R_{\sst{V}}^{(8)}}}
\def\raz{{R_{\sst{A}}^{(0)}}}
\def\rat{{R_{\sst{A}}^{(3)}}}
\def\rae{{R_{\sst{A}}^{(8)}}}
\def\rvtez{{R_{\sst{V}}^{T=0}}}
\def\rvteo{{R_{\sst{V}}^{T=1}}}
\def\ratez{{R_{\sst{A}}^{T=0}}}
\def\rateo{{R_{\sst{A}}^{T=1}}}
\def\mro{{m_\rho}}
\def\mks{{m_{\sst{K}}^2}}
\def\mpi{{m_\pi}}
\def\mpis{{m_\pi^2}}
\def\mom{{m_\omega}}
\def\mphi{{m_\phi}}
\def\Qhat{{\hat Q}}
\def\FOS{{F_1^{(s)}}}
\def\FTS{{F_2^{(s)}}}
\def\GAS{{G_{\sst{A}}^{(s)}}}
\def\GES{{G_{\sst{E}}^{(s)}}}
\def\GMS{{G_{\sst{M}}^{(s)}}}
\def\GATEZ{{G_{\sst{A}}^{\sst{T}=0}}}
\def\GATEO{{G_{\sst{A}}^{\sst{T}=1}}}
\def\mdax{{M_{\sst{A}}}}
\def\mustr{{\mu_s}}
\def\rsstr{{r^2_s}}
\def\rhostr{{\rho_s}}
\def\GEG{{G_{\sst{E}}^\gamma}}
\def\GEZ{{G_{\sst{E}}^\sst{Z}}}
\def\GMG{{G_{\sst{M}}^\gamma}}
\def\GMZ{{G_{\sst{M}}^\sst{Z}}}
\def\GEn{{G_{\sst{E}}^n}}
\def\GEp{{G_{\sst{E}}^p}}
\def\GMn{{G_{\sst{M}}^n}}
\def\GMp{{G_{\sst{M}}^p}}
\def\GAp{{G_{\sst{A}}^p}}
\def\GAn{{G_{\sst{A}}^n}}
\def\GA{{G_{\sst{A}}}}
\def\GETEZ{{G_{\sst{E}}^{\sst{T}=0}}}
\def\GETEO{{G_{\sst{E}}^{\sst{T}=1}}}
\def\GMTEZ{{G_{\sst{M}}^{\sst{T}=0}}}
\def\GMTEO{{G_{\sst{M}}^{\sst{T}=1}}}
\def\lamd{{\lambda_{\sst{D}}^\sst{V}}}
\def\lamn{{\lambda_n}}
\def\lams{{\lambda_{\sst{E}}^{(s)}}}
\def\bvz{{\beta_{\sst{V}}^0}}
\def\bvo{{\beta_{\sst{V}}^1}}
\def\Gdip{{G_{\sst{D}}^\sst{V}}}
\def\GdipA{{G_{\sst{D}}^\sst{A}}}
\def\fks{{F_{\sst{K}}^{(s)}}}
\def\FIS{{F_i^{(s)}}}
\def\fpi{{F_\pi}}
\def\fk{{F_{\sst{K}}}}
\def\RAp{{R_{\sst{A}}^p}}
\def\RAn{{R_{\sst{A}}^n}}
\def\RVp{{R_{\sst{V}}^p}}
\def\RVn{{R_{\sst{V}}^n}}
\def\rva{{R_{\sst{V,A}}}}
\def\xbb{{x_B}}
\def\mlq{{M_{\sst{LQ}}}}
\def\mlqs{{M_{\sst{LQ}}^2}}
\def\lscal{{\lambda_{\sst{S}}}}
\def\lvect{{\lambda_{\sst{V}}}}
\def\PR#1{{{\em   Phys. Rev.} {\bf #1} }}
\def\PRC#1{{{\em   Phys. Rev.} {\bf C#1} }}
\def\PRD#1{{{\em   Phys. Rev.} {\bf D#1} }}
\def\PRL#1{{{\em   Phys. Rev. Lett.} {\bf #1} }}
\def\NPA#1{{{\em   Nucl. Phys.} {\bf A#1} }}
\def\NPB#1{{{\em   Nucl. Phys.} {\bf B#1} }}
\def\AoP#1{{{\em   Ann. of Phys.} {\bf #1} }}
\def\PRp#1{{{\em   Phys. Reports} {\bf #1} }}
\def\PLB#1{{{\em   Phys. Lett.} {\bf B#1} }}
\def\ZPA#1{{{\em   Z. f\"ur Phys.} {\bf A#1} }}
\def\ZPC#1{{{\em   Z. f\"ur Phys.} {\bf C#1} }}
\def\etal{{{\em   et al.}}}
\def\delalr{{{delta\alr\over\alr}}}
\def\pbar{{\bar{p}}}
\def\lamchi{{\Lambda_\chi}}
\def\qw0{{Q_{\sst{W}}^0}}
\def\qwp{{Q_{\sst{W}}^P}}
\def\qwn{{Q_{\sst{W}}^N}}
\def\qwe{{Q_{\sst{W}}^e}}
\def\qem{{Q_{\sst{EM}}}}
\def\gae{{g_{\sst{A}}^e}}
\def\gve{{g_{\sst{V}}^e}}
\def\gvf{{g_{\sst{V}}^f}}
\def\gaf{{g_{\sst{A}}^f}}
\def\gvu{{g_{\sst{V}}^u}}
\def\gau{{g_{\sst{A}}^u}}
\def\gvd{{g_{\sst{V}}^d}}
\def\gad{{g_{\sst{A}}^d}}
\def\gvftil{{\tilde g_{\sst{V}}^f}}
\def\gaftil{{\tilde g_{\sst{A}}^f}}
\def\gvetil{{\tilde g_{\sst{V}}^e}}
\def\gaetil{{\tilde g_{\sst{A}}^e}}
\def\gvqtil{{\tilde g_{\sst{V}}^e}}
\def\gaqtil{{\tilde g_{\sst{A}}^e}}
\def\gvutil{{\tilde g_{\sst{V}}^e}}
\def\gautil{{\tilde g_{\sst{A}}^e}}
\def\gvdtil{{\tilde g_{\sst{V}}^e}}
\def\gadtil{{\tilde g_{\sst{A}}^e}}
\def\delp{{\delta_P}}
\def\delzp{{\delta_{00}}}
\def\deld{{\delta_\Delta}}
\def\dele{{\delta_e}}
\def\lnew{{{\cal L}_{\sst{NEW}}}}
\def\osffp{{{\cal O}_{7a}^{ff'}}}
\def\oszg{{{\cal O}_{7c}^{Z\gamma}}}
\def\osgg{{{\cal O}_{7b}^{g\gamma}}}
\def\slash#1{#1\!\!\!{/}}
\def\beq{\begin{eqnarray}}
\def\eeq{\end{eqnarray}}
\def\bea{\begin{eqnarray*}}
\def\eea{\end{eqnarray*}}
\def\NCA{\em Nuovo~Cimento}
\def\IJMP{\em Intl.~J.~Mod.~Phys.}
\def\NP{\em Nucl.~Phys.}
\def\PLB{{\em Phys.~Lett.}~B}
\def\JETPLett{{\em JETP Lett.}}
\def\PRL{\em Phys.~Rev.~Lett.}
\def\MPL{\em Mod.~Phys.~Lett.}
\def\PRD{{\em Phys.~Rev.}~D}
\def\PR{\em Phys.~Rev.}
\def\PRP{\em Phys.~Rep.}
\def\ZPC{{\em Z.~Phys.}~C}
\def\PTP{{\em Prog.~Theor.~Phys.}}
% Some other macros used in the sample text
\def\Baryon{{\rm B}}
\def\Lepton{{\rm L}}
\def\sbar{\overline}
\def\stilde{\widetilde}
\def\st{\scriptstyle}
\def\sst{\scriptscriptstyle}
\def\vac{|0\rangle}
\def\argh{{{\rm arg}}}
\def\G{\stilde G}
\def\Wmess{W_{\rm mess}}
\def\NI{\stilde N_1}
\def\antivac{\langle 0|}
\def\infinity{\infty}
\def\mco{\multicolumn}
\def\epp{\epsilon^{\prime}}
\def\psibar{\overline\psi}
\def\nmess{N_5}
\def\chibar{\overline\chi}
\def\lagr{{\cal L}}
\def\drbar{\overline{\rm DR}}
\def\msbar{\overline{\rm MS}}
\def\conj{{{\rm c.c.}}}
\def\Et{{\slashchar{E}_T}}
\def\Etot{{\slashchar{E}}}
\def\mZ{m_Z}
\def\MPlanck{M_{\rm P}}
\def\mW{m_W}
\def\cbeta{c_{\beta}}
\def\sbeta{s_{\beta}}
\def\cW{c_{W}}
\def\sW{s_{W}}
\def\deltaeps{\delta}
\def\sigmabar{\overline\sigma}
\def\epsilonbar{\overline\epsilon}
\def\vep{\varepsilon}
\def\ra{\rightarrow}
\def\half{{1\over 2}}
\def\ko{K^0}
\def\be{\beq}
\def\ee{\eeq}
\def\bea{\begin{eqnarray}}
\def\eea{\end{eqnarray}}

%  \gsim and \lsim provide >= and <= signs.
\def\centeron#1#2{{\setbox0=\hbox{#1}\setbox1=\hbox{#2}\ifdim
\wd1>\wd0\kern.5\wd1\kern-.5\wd0\fi
\copy0\kern-.5\wd0\kern-.5\wd1\copy1\ifdim\wd0>\wd1
\kern.5\wd0\kern-.5\wd1\fi}}
\def\ltap{\;\centeron{\raise.35ex\hbox{$<$}}{\lower.65ex\hbox{$\sim$}}\;}
\def\gtap{\;\centeron{\raise.35ex\hbox{$>$}}{\lower.65ex\hbox{$\sim$}}\;}
\def\gsim{\mathrel{\gtap}}
\def\lsim{\mathrel{\ltap}}

%%%%%%%%%%%%%%%%%%%%%%%%%%%%%%%%%%%%%%%
%  Slash character...
\def\slashchar#1{\setbox0=\hbox{$#1$}           % set a box for #1
   \dimen0=\wd0                                 % and get its size
   \setbox1=\hbox{/} \dimen1=\wd1               % get size of /
   \ifdim\dimen0>\dimen1                        % #1 is bigger
      \rlap{\hbox to \dimen0{\hfil/\hfil}}      % so center / in box
      #1                                        % and print #1
   \else                                        % / is bigger
      \rlap{\hbox to \dimen1{\hfil$#1$\hfil}}   % so center #1
      /                                         % and print /
   \fi}                                        %

%%EXAMPLE:  $\slashchar{E}$ or $\slashchar{E}_{t}$
\setcounter{tocdepth}{2}

%%%%%%%%%%%%%%%%%%%%%%%%%%%%%%%%%%%%%%%%%%%%%%%%%%%

\title{Radiative corrections to low energy neutrino reactions}
\author {A. Kurylov$^{a}$, M.J. Ramsey-Musolf$^{a,b,c}$, and P. Vogel$^{a}$ }
\address{$^{a}$Kellogg Radiation Laboratory and Physics Department \\
Caltech, Pasadena, CA 91125 \\
$^b$ Department of Physics, University of Connecticut, Storrs, CT  06269\\
$^c$ Institute for Nuclear Theory, University of Washington, Seattle, WA 98195}
\date{\today}
\maketitle

\begin{abstract}
We show that the radiative corrections to charged current (CC) nuclear
reactions with an
electron(positron) in the final state are described by a universal function.
The consistency of our treatment of the radiative corrections with the
procedure
used to extract the value of the axial coupling constant $g_A$ is discussed.
To illustrate we apply our results to (anti)neutrino deuterium
disintegration and to $pp$ fusion in the sun.
The limit of vanishing electron mass is considered, and a simple formula
valid for $E_{obs}\gsim 1$ MeV is obtained.
The size of the
nuclear structure-dependent effects is also discussed.
Finally, we
consider CC transitions with an electron(positron) in the initial state and
discuss some applications to electron capture reactions.
\\
\end{abstract}

PACS number(s): 13.10.+q, 13.15.+g, 25.30.Pt

\vspace{1cm}
%%%%%%%%%%%%%%%%%%%%%%%%%%%%%%%%%%%%%%

\section{Introduction}
\label{sec:intro}

The physics of neutrino flavor oscillations has entered a new era.
There is now a consensus that the observation
of the atmospheric neutrinos \cite{SK} can be interpreted
as an evidence that muon neutrinos born in the atmosphere oscillate
into tau neutrinos.
At the same time, numerous measurements of solar electron
neutrino fluxes have been pointing toward neutrino
oscillations ever since the Homestake experiment \cite{homestake}.
The evidence for oscillation of solar neutrinos has been strengthened recently
with the results from Sudbury Neutrino Observatory (SNO) \cite{SNO,SNO-NC} that
are independent of the solar model (SSM).
With the evidence for oscillations at hand, the accurate determination of
the neutrino masses and flavor mixing angles
is the most important task for neutrino physics at present.
Carrying out the above program requires accurate
knowledge of the cross sections of the reactions used for
neutrino detection, and the Standard
Model (SM) radiative corrections,
which typically shift the leading  (tree)
order values by 3-4\%, must be taken into account.

Complete one-loop SM radiative corrections to the CC and neutral current (NC)
deuteron disintegration by electron neutrinos
used in SNO measurements have been
calculated in Refs. \cite{krmv01,Towner}.  We have shown in Ref. \cite{krmv01}
that while the differential
CC cross section depends on the actual detector
properties ({\it e.g.} on the bremsstrahlung
detection threshold $E_\gamma^{min}$) the total cross section
is a detector-independent quantity as long as
the final state electron is always detected, i.e., the total number
of the neutrino interactions is determined.
In the simple case when $E_\gamma^{min}\rightarrow 0$ one finds:
\beq
\label{eq:g-zero-thresh}
d\sigma_{CC}(E_{obs})=d\sigma^{\rm Tree}_{CC}(E_{obs})
\left(1+{\alpha\over \pi} g(E_{obs}) \right) ~,
\eeq
where $d\sigma^{\rm Tree}_{CC}(E_{obs})$ and $d\sigma_{CC}(E_{obs})$ are,
respectively, the leading and
the next to leading order in $\alpha$
differential cross sections, $E_{obs}$ is the energy observed in the detector
(charged lepton energy plus possible bremsstrahlung photon energy),
and $g(E_{obs})$ is given by the Eq. (\ref{eq:g-total}) below.
It is understood that the cross section
$d\sigma^{\rm Tree}_{CC}(E_{obs})$ includes the Fermi function
$F(Z,A,E_{obs})$
to account for the
distortion of the final state electron wavefunction by the
Coulomb field of the final nucleus (see {\it e.g.}
Ref. \cite{preston}). While a similar approach to the treatment of
radiative corrections has long been used in the analysis of nuclear beta decay
\cite{Sirlin-g,Sirlin}, here we extend it to a wide class of reactions
involving neutrinos, and formulate the conditions for its applicability
\footnote{
The connection between $g(E)$ and the function $G(E_m,E)$ introduced by Sirlin
\cite{Sirlin-g} for the description of allowed $\beta$-decays is
discussed in Section \ref{sec:mass-singular}.}.
Specifically, we generalize the results of Ref. \cite{krmv01} and analyze
the following four features
of the radiative corrections.

1. \underline {Universality}. The function $g(E)$ is {\it universal}
for a class of nuclear reactions involving neutrinos.
In particular, it
can be used to evaluate the radiative corrections to the total cross-sections
of reactions that have an electron (positron) in the final state.
It is applicable not only to neutron and nuclear $\beta$-decays,
but also to the antineutrino capture reactions, low energy CC neutrino-nucleus
disintegration, nuclear fusion reactions accompanied by
the electron(positron) emission, {\it etc.}
For illustration we consider the following CC processes
involving two nucleon systems:
\beqa
\label{eq:reactions}
\nu_e+d &\rightarrow& p+p+e^- ~, \nonumber \\
{\bar \nu}_e+d &\rightarrow& n+n+e^+ ~, \nonumber \\
p+p &\rightarrow& d+e^+ +  \nu_e ~.
\eeqa
For reactions where the electron spectrum is narrow (precise conditions are
given in
Section \ref{sec:universality})
we provide a simple and accurate way to calculate the correction to the
total cross section
without the complicated integration over the outgoing electron or positron
spectrum.

In order to demonstrate
one of the possible applications of our results, note that
the last reaction in Eq. (\ref{eq:reactions}) is the main reaction for the
$pp$ chain
that powers the
Sun (see {\it e.g.} Ref. \cite{bahcall-review}). The total rate of this
reaction is,
therefore, constrained by the known solar luminosity.
That rate, together with the cross-section for the
$pp$ reaction, is used to calibrate the SSM and
to predict many other quantities, in particular
the flux of the $^8$B neutrinos studied at SNO.
Below we show that a proper treatment of the radiative corrections
to $S_{pp}$, the S-factor for the $pp$ reaction, lowers the SSM prediction
to $\Phi(^8{\rm B})$
by about 0.6\% relative to the currently accepted value \cite{solar-x-sect}.
\footnote{
The \lq\lq outer" radiative corrections
to $S_{pp}$ have been originally evaluated in Ref.\cite{pp-orig},
whose results are, however, not used in Ref. \cite{solar-x-sect}.
Results obtained using our simple prescription agree with the
calculation presented in Ref. \cite{pp-orig}.
}

2. \underline{Axial coupling constant $g_A$}.
Since the reactions in Eq.(\ref{eq:reactions}) are dominated by
the nuclear axial current, it is important to use a
procedure consistent with the definition
of the axial coupling constant $g_A$.
We briefly discuss the conventional definition of $g_A$ based
on the value extracted from neutron decay and give a prescription
for evaluation of the radiative corrections
consistent with that definition. This issue has been discussed in the
literature (see {\it e.g. } Ref. \cite{garcia})
but we feel that it is sufficiently important to reiterate here.

3. \underline{Nuclear structure-dependent effects}.
In deriving universality properties of the radiative corrections we work in
the \lq\lq one-body" approximation (in analogy to the early applications
to the nuclear beta decay). Namely, we evaluate the radiative
correction to the CC reaction on a single
nucleon, and then use it to compute the correction
to the reaction of interest. Effectively,
only the corrections of the type shown in Fig.\ref{fig:rc-types}(a)
are included in such a calculation.
Nuclear structure-dependent many-body corrections of the type shown in
Fig.\ref{fig:rc-types}(b)
are thus neglected. We show that a comparison of such corrections to
analogous effects contributing to
superallowed Fermi transitions in nuclei suggests that they enter at 0.1\%
level. As we
discuss in Section \ref{sec:manu-nucl}, however, this estimate of
universality-breaking
contributions is not airtight,
and a detailed computation is needed to obtain precise numbers.

4. \underline{Collinear singularities}.
We study the behavior of the radiative corrections in the limit
$m_e\rightarrow 0$,
where $m_e$ is the electron mass. Separate contributions to the corrections
are singular in this limit, and until now it has not been explicitly shown
in the literature how
the singularities cancel when all contributions are added. Such cancellation
must take place according to the well-known Kinoshita-Lee-Nauenberg (KLN)
theorem\cite{kln},
and in Section \ref{sec:mass-singular} we show how this happens.
As a corollary we obtain a simple expression for $g(E)$ valid for $E\gg
m_e$. In practice, radiative
corrections evaluated according to this simple expression are accurate to
about 0.1\% for energies above
1 MeV.

Although these four topics constitute the main focus of our study, we also
discuss processes, such
as capture reactions, involving charged leptons in the initial state. In
this case, the radiative
corrections do not display the same kind of universality applicable to the
reactions in Eq.
(\ref{eq:reactions}). Nevertheless, the treatment of the two cases is
sufficiently similar that we
feel a brief discussion -- given at the end of the paper -- is warranted.

\section{Universality of the radiative corrections to
reactions with the electron/positron in the final state}
\label{sec:universality}

We begin with the first reaction in Eq.(\ref{eq:reactions}).
As was shown in Ref. \cite{krmv01}, the total cross section is independent
of the photon detection
threshold $E_\gamma^{min}$. In the limit $E_\gamma^{min}\to 0$ the function
$g(E)$
in Eq.(\ref{eq:g-zero-thresh}) has two parts:
\beq
\label{eq:g-total}
g(E)=g_v(E)+g_b(E) ~.
\eeq
Here, $g_v(E)$ (for the virtual part)
and $g_b(E)$ (for the bremsstrahlung part)
are given by the following lengthy expressions (the cutoff parameter
$\Lambda$ that appears in Ref. \cite{krmv01,Towner}
is set equal to the proton mass):
\beqa
\label{eq:gv}
g_v(E)&=&2\ln\left( {M_Z\over M_p} \right)
+\frac{3}{2} ~ {\rm ln}  \left( \frac{ M_p } { m_e } \right) \nonumber \\
&+&2\ln\left( {E-m_e\over m_e} \right)
\left( {1\over 2 \beta(E)} \ln \left( {1+\beta(E) \over 1-\beta(E) }
\right)-1\right) \nonumber \\
&+&{3\over 4} +{\cal A}(\beta(E))-0.57~, \nonumber \\
{\cal A}(\beta)&=& \frac{1}{2} \beta ~ {\rm ln} \left( \frac{1 + \beta}{1 -
\beta} \right) -1
- \frac{1}{ \beta }
\left[ \frac{1}{2} {\rm ln} \left( \frac{1 + \beta}{1 - \beta} \right)
\right]^2
+ \frac{1}{ \beta } L \left( \frac{ 2 \beta }{ 1 + \beta } \right)~.
\nonumber \\
\eeqa
and
\beqa
\label{eq:gb}
g_b(E)&=&{\cal C}(\beta(E))+{1\over 2 E^2 \beta(E) }\Biggl(
\int_{m_e}^E (E-x)\ln\left( {1+\beta(x) \over 1-\beta(x)} \right)dx
\nonumber \\
&+&4E \int_{m_e}^E { x\beta(x)F(x)-E\beta(E)F(E) \over E-x}dx \Biggr)~,
\nonumber \\
F(E)&=&{1\over 2\beta(E)}\ln\left({1+\beta(E) \over 1-\beta(E)}\right)-1~,
\nonumber \\
{\cal C}(\beta)&=&2\ln(2)\left[
\frac{1}{2 \beta}  \ln  \left( \frac{1 + \beta}{1 - \beta} \right) -1
\right]+1+\frac{1}{4 \beta}  \ln  \left( \frac{1 + \beta}{1 - \beta}
\right) \nonumber \\
& & \times \left[ 2+ \ln \left( {{1-\beta^2} \over 4}
\right) \right] + {1 \over {\beta}} \left[ {L(\beta)-L(-\beta)} \right]
\nonumber \\
&+&{1 \over 2 \beta} \left( {L\left( {{1-\beta} \over {2}} \right) - L\left(
{{1+\beta} \over {2}} \right)} \right)~.
\eeqa
In the above equations, $m_e$ is the electron mass and:
\beqa
\label{eq:functions}
\beta(E)&=&{\sqrt{E^2-m_e^2}\over |E|}~, \nonumber \\
L(\beta)& = & \int_0^{\beta}{{\ln(| 1-x |)}\over x}dx ~
{\stackrel {\strut |\beta|\leq 1}{=}} ~ -
\sum_{k=1}^{\infty}\frac{\beta^k}{k^2} ~.
\eeqa
The closed form for the integrals appearing in Eq. (\ref{eq:gb}) is given in the appendix.
Although $g(E_{obs})$ in general depends on the value of the cutoff
parameter $\Lambda$,
the choice of $\Lambda$ affects only the constant, energy independent part
of $g_v(E_{obs})$. We chose $\Lambda = M_p$ and adjusted the constant (-0.57)
in the Eq.(\ref{eq:gv}) so that the result matches with the corresponding
expression derived by Sirlin in Ref. \cite{Sirlin} based on current algebra
(the same approach was chosen in Ref. \cite{Towner}). The dominant
uncertainty in $g(E_{obs})$
is associated with the value of the matching constant and is briefly
discussed after Eq.
(\ref{eq:g-low}).

It is remarkable that $g(E_{obs})$ is a function of a single parameter --
the observed
energy $E_{obs}$ -- and does not depend on the electron and photon
energies separately. Moreover,
formulas Eq.(\ref{eq:gv}) and Eq.(\ref{eq:gb}) make no reference
to the parameters that describe the
leading-order differential cross section, such as $pp$ scattering length
or the effective range
(see {\it e.g.} Ref. \cite{Kelly} for definitions).
It is, therefore, reasonable to ask whether $g(E_{obs})$
is a universal function that describes the radiative corrections to a whole
class
of processes in the limit $E_\gamma^{min}\rightarrow 0$.
Below we show that this is indeed the case.

Consider first, in the one-nucleon approximation,
contributions from virtual photon exchanges. As
pointed out in Refs. \cite{krmv01,Towner}, it is convenient to
split these contribution into two pieces: with
$\kappa<\Lambda$ and with $\kappa\ge\Lambda$,
where $\kappa$ is the scale for the virtual photon momenta and
$\Lambda$ is a cutoff, taken to be of the order of 1 GeV.
Contributions with $\kappa\ge\Lambda$,
usually called \lq\lq inner" radiative corrections, are combined with the
$Z^0$ boson exchange
box graphs to produce an energy-independent contribution to $g(E_{obs})$
\cite{Towner}:
\beq
\label{eq:outer}
g_v(E_{obs})^{\kappa\ge\Lambda}=2\ln\left(M_Z\over\Lambda\right) ~.
\eeq
This contribution is obviously common to all reactions in
Eq.(\ref{eq:reactions}) since
it contains no parameters that distinguish between them.

\begin{figure}
\begin{center}
\includegraphics[width=5in]{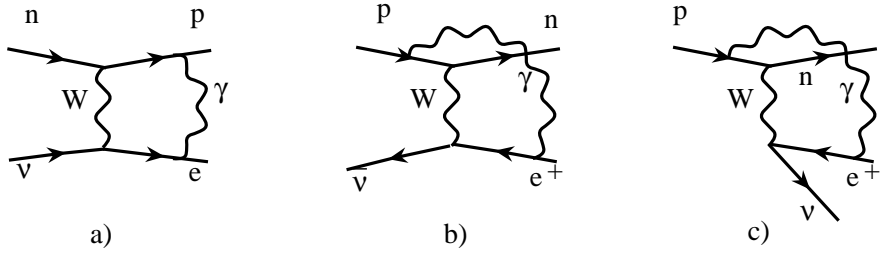}
\caption{Exchanges of a low energy virtual photon for the processes shown
in Eq.(\ref{eq:reactions}) in the
one-nucleon approximation.}
\label{fig:virt-exch}
\end{center}
\end{figure}

The piece with $\kappa<\Lambda$ is more complicated.
The virtual exchanges of a low energy photon
for all three reactions Eq.(\ref{eq:reactions}) are shown in
Fig.\ref{fig:virt-exch}.
The correction to
Fig.\ref{fig:virt-exch}(a) is \cite{krmv01,Towner}:
\beqa
\label{eq:g-low}
g_v(E_{obs})^{\kappa<\Lambda}&=&3\ln\left({\Lambda\over M_p}\right)
+{3\over 2}\ln\left({M_p\over m_e}\right)+{3\over 4} +{\cal
A}(\beta(E_{obs})) \nonumber \\
&+&2\ln\left({\lambda\over m_e}\right)\left[{1\over 2\beta(E_{obs})}
\ln\left({1+\beta(E_{obs})\over
1-\beta(E_{obs})}\right)-1\right]-0.57+\delta{\stilde C}(\Lambda)~,
\eeqa
where $\lambda$ is the \lq\lq photon mass" used as the infrared regulator and
${\cal A}(\beta)$ is given in Eq. (\ref{eq:gv}). The parameter
$\delta{\stilde C}$ represents a
matching constant arising from presently incalculable non-perturbative QCD
effects. Its
$\Lambda$-dependence must cancel the corresponding
dependence in $g(E_{obs})$. Unfortunately, no existing calculation of
$\delta{\stilde C}$ has
demonstrated this cancellation explicitly. In addition, $\delta{\stilde C}$
will contain
$\Lambda$-independent terms that depend on the light-quark masses,
$\Lambda_{\rm QCD}$, {\em etc.}.
A complete, first principles computation of these contributions has yet to
be performed, so
we must rely on models. Thus,
in practical calculations we set this constant to zero for $\Lambda=M_p$ to
match the model results of Refs. \cite{Sirlin,towner-c,towner-new}. The
uncertainty in $g_v(E)$
associated with $\delta{\stilde C}$ is estimated in Ref. \cite{towner-new}
to be about 0.08\%.

Even though the virtual momenta in Figs.\ref{fig:virt-exch}(b,c)
are different from those
on Fig.\ref{fig:virt-exch}(a) their contributions are also given
by the Eq.(\ref{eq:g-low}). The easiest way to
see it is as follows. Consider Fig.\ref{fig:virt-exch}(b) first.
This graph can be obtained from
Fig. \ref{fig:virt-exch}(a) through time reversal followed by replacements
$e^-({\rm initial})\rightarrow e^+({\rm final})$ and
$\nu_e({\rm final})\rightarrow {\bar \nu}_e({\rm initial})$.
Since time reversal
is an exact symmetry of QED, the scattering amplitude
is unchanged up to an overall phase.
The result of replacements of the particles with antiparticles
can be found by utilizing  crossing symmetry.
However, the replacement $p_e\rightarrow -p_e$
does not change $\beta(E_e)$ (see Eq.(\ref{eq:functions})
for the definition). Since Eq.
(\ref{eq:g-low}) depends on the electron momenta only through $\beta(E_e)$
it is obviously
invariant.
Hence, Eq. (\ref{eq:g-low}) is applicable to
the second process in Eq.(\ref{eq:reactions}).
The validity of the Eq. (\ref{eq:g-low}) for the third process in
Eq.(\ref{eq:reactions})
trivially follows from the fact that Eq. (\ref{eq:g-low})
is independent of the neutrino 4-momentum. Indeed, going
from Fig.\ref{fig:virt-exch}(b) to Fig.\ref{fig:virt-exch}(c),
which is accomplished by replacing
the antineutrino with the neutrino and changing the sign of its 4-momentum,
leaves the Eq. (\ref{eq:g-low}) invariant.

\begin{figure}
\begin{center}
\includegraphics[width=2in]{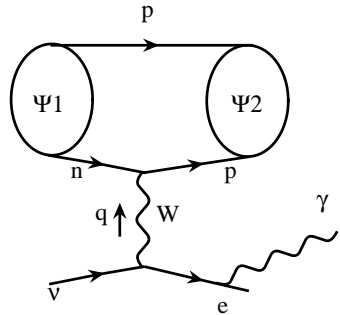}
\caption{The dominant bremsstrahlung graph. $\Psi_1$ and $\Psi_2$ represent
the initial and the final
nuclear wavefunctions, respectively. $q$ is the 4-momentum transfer from
the leptonic to the hadronic
part of the graph. In general, $\Psi_{1,2}$ can contain any number of
nucleons.}
\label{fig:bremsst}
\end{center}
\end{figure}

The bremsstrahlung contributions are also identical for all
three reactions under consideration [Eq.(\ref{eq:reactions})]. In
particular, for these reactions,
only the bremsstrahlung from the charged lepton is important since all other
charged particles are heavy. Therefore, for each of the
processes under consideration there is only one
bremsstrahlung diagram of the type shown in Fig.\ref{fig:bremsst}.
The amplitude for this graph has the form:
\beq
\label{eq:mat-el-gamma}
{\cal M}_\gamma\sim{e G_F\over \sqrt 2} \epsilon_\nu L_\gamma^{\mu\nu} 
H_\mu(q) ~,
\eeq
where $e$ is the electron charge, $\epsilon_\nu$ is the photon
polarization, and
$L_\gamma^{\mu\nu}$ and $H_\mu(q)$ represent, respectively,
the leptonic and the hadronic contributions to the process:
\beqa
H_\mu(q)&=&\int e^{iq\cdot x} \langle f_N |\sum_{m=1}^A J_\mu^m(x)| i_N
\rangle d^4x \nonumber \\
L_\gamma^{\mu\nu}&=&\int e^{-iq\cdot x-ik\cdot y}
\langle f_L | T\{J^\mu(x) J_{\rm EM}^\nu(y) \}| i_L \rangle d^4xd^4y ~.
\eeqa
Here $i_{N,L}$ and $f_{N,L}$ are, respectively,
the initial and the final states of the nuclear and the leptonic
parts of the process, $J_\mu$ is the relevant charged current,
$J_{\rm EM}^\nu$ is the electromagnetic
current, and $k$ is the momentum of the emitted photon.
The summation index $m$ in the hadronic matrix element runs over
all nucleons in the nucleus.
The dependence of $H_\mu$ on the momentum transfer $q$
is explicitly shown. In the long wavelength approximation, valid
for low neutrino energies,
one can neglect the dependence of $H_\mu$ on the spatial components of $q$.
On the other hand, the dependence of $H_\mu$ on the time
component of $q$ could be crucial.
Indeed, energy conservation implies that
the initial $E_i$ and  final $E_f$ energies of the hadronic system obey the
relation
$E_f=E_i+q^0$. Clearly, the nuclear matrix element depends sensitively
on the value of $E_f$, the excitation energy of the final state.
In the case of CC neutrino deuteron disintegration, for example,
changing $E_f$ causes a dramatic
change in the shape of the final state $pp$ wavefunction and the corresponding
change in the size of the nuclear matrix element\cite{krmv01}.

With these consideration in mind, and using the results of
Ref. \cite{krmv01,Towner}, it is easy to show that the
square of the matrix element Eq.(\ref{eq:mat-el-gamma})
averaged over the initial and summed over the final spins has the form
\beqa
\label{eq:mat-elem}
\langle |{\cal M}_\gamma(E_\nu,E_e,E_\gamma,x)|^2 \rangle&=&{\alpha\over \pi}
\langle |{\cal M}_0(E_\nu,E_{obs})|^2 \rangle \times
{E_e\over E_{obs}}F(E_\gamma,E_e,x) ~, \nonumber \\
q^0&\equiv&\pm E_\nu-E_e-E_{\gamma} = \pm E_\nu-E_{obs} ~.
\eeqa
Here $\langle |{\cal M}_0(E_\nu,E_{obs})|^2 \rangle$ is
the spin-averaged square of the matrix element
in the leading order of perturbation theory and
$x$ is the cosine of the angle between the momenta of the photon and
the electron.
The dependence of ${\cal M}_0$ on $q^0$ is implicit through the second
line of Eq.(\ref{eq:mat-elem}),
where the sign of the neutrino energy depends on
whether the neutrino is in the initial (+) or final (-) state.
$F(E_\gamma,E_e,x)$ is given by the second line
of Equation (11) in Ref. \cite{Towner}
\footnote{Our definition differs from that in Ref. \cite{Towner} by a
factor of 2, which if absorbed in
$F(E_\gamma, E_e, x)$ to simplify the expression in Eq. (\ref{eq:mat-elem}).
}:
\beqa
\label{eq:F-func}
F(E_\gamma,E_e,x)&=&{E_\gamma\over{2E_e^2(E_\gamma-\beta(E_e)Kx)}}
+\left(E_e+E_\gamma\over 2E_e \right)
{\beta^2(E)(1-K^2x^2/E_\gamma^2)\over (E_\gamma-\beta(E_e)Kx)^2}~,
\nonumber \\
K^2&=&E_\gamma^2-\lambda^2 ~.
\eeqa
As before, $\lambda$ is the infrared regulator (``photon mass'').
While the form of the matrix element ${\cal M}_0$  may depend
on the particular process under consideration,
the function $F(E_\gamma,E_e,x)$ is universal.

In order to complete the calculation of the bremsstrahlung
contribution to the cross section
it is sufficient to multiply the first line of Eq.(\ref{eq:mat-elem})
by the appropriate phase space factor
and integrate over the final state momenta to obtain
\beqa
\label{eq:br-cross}
d\sigma_{CC}^\gamma=d\sigma^{\rm Tree}_{CC}(E_\nu,E_{obs})\times
{\alpha\over \pi}
{\beta(E_e)E_e^2\over{\beta(E_{obs})E_{obs}^2}}F(E_\gamma,E_e,x)dxK{dKdE_e\over
dE_{obs}} ~,
\eeqa
where again $E_{obs}=E_e+E_{\gamma}$ and $d\sigma^{\rm
Tree}_{CC}(E_\nu,E_{obs})$
is the leading order differential cross section.
If there is at least one more particle in the final
state except the electron and the photon, its phase space factor can be
used to eliminate
the spatial part of the overall momentum-conserving delta function, which
is always
present in the expression for the differential cross section. Keeping
$E_{obs}$ fixed, one can then perform integrations over $x$ and $E_e$.
Changing the integration variables from $\{E_e,K\}$ to
$\{E_e,E_{obs}\}$ and integrating over $E_e$ from $m_e$ to $E_{obs}$ we obtain:
\beqa
\label{eq:brem-int}
d\sigma^\gamma_{CC}(E_\nu,E_{obs})
&=&d\sigma^{\rm Tree}_{CC}(E_\nu,E_{obs}) \times {\alpha\over \pi}
\int_{m_e}^{E_{obs}}dE_e \int_{-1}^1dx \nonumber \\
&\times&{\beta(E_e)E_e^2\over{\beta(E_{obs})E_{obs}^2}}F(E_{obs}-E_e,E_e,x)
(E_{obs}-E_e) \nonumber \\
&\equiv&d\sigma^{\rm Tree}_{CC}(E_\nu,E_{obs}) \times {\alpha\over \pi}
g_{brem}(E_{obs}) ~.
\eeqa
The above integration is delicate due to the presence of infrared divergence.
The result, however, is well
known (see {\it e.g.} Ref. \cite{Towner}):
the infrared divergence in $g_{brem}$ cancels that in Eq.(\ref{eq:g-low}).

\begin{figure}
\begin{center}
\includegraphics[width=4in]{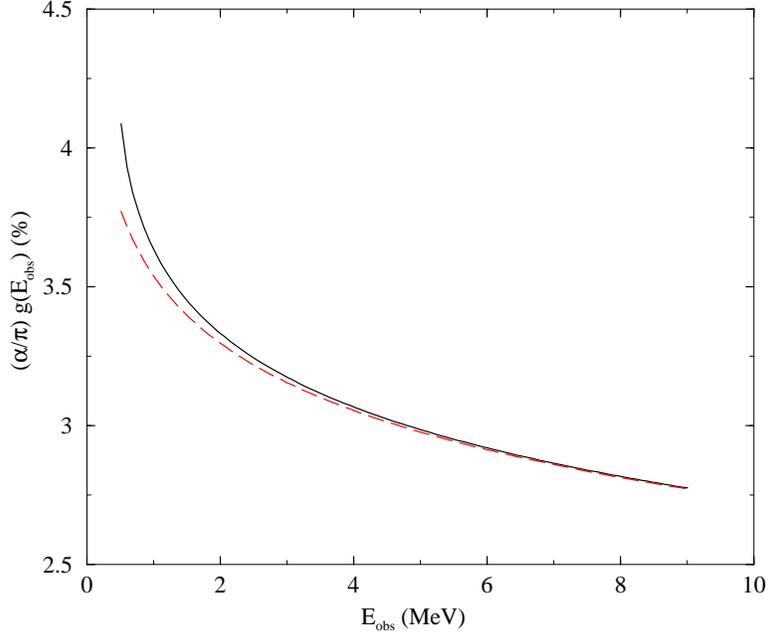}
\caption{The exact one-loop radiative correction ${(\alpha/\pi)}g(E_{obs})$
in \% (solid line) for reactions
in Eq.(\ref{eq:reactions}) and the
same correction in the limit $m_e\to 0$ (dashed line).}
\label{fig:g-function}
\end{center}
\end{figure}

Now, the final result for the sum of all corrections is
\beq
\label{eq:genresult}
g(E_{obs})=g_v(E_{obs})^{\kappa\ge\Lambda}+g_v(E_{obs})^{\kappa\le\Lambda}+g_{brem}(
E_{obs}) ~,
\eeq
with $g(E)$ given in Eq.(\ref{eq:g-total}) for $\Lambda=M_p$.
This completes the proof of the universality
of this function. The dependence of $g(E)$ on energy is shown in
Fig.\ref{fig:g-function}.

\section{Examples}
\label{sec:examples}

With help of Eq.(\ref{eq:g-zero-thresh}), the function $g(E_{obs})$
obtained in the
previous section can be used to account for the radiative corrections to
the differential cross section
to all reactions listed in Eq.(\ref{eq:reactions}). A dramatic
simplification occurs if one is only
interested in computing the total cross section.
The latter is given by the integral
\beq
\label{eq:exact}
\sigma_{CC}=\int_{m_e}^{E^{max}} d\sigma^{\rm Tree}_{CC}(E)
\left(1+{\alpha\over \pi} g(E) \right)
\eeq
where the subscript \lq\lq obs" on $E$ is not shown. We argue below that
if the leading order electron spectrum is sufficiently narrow there is no need to
compute the integral explicitly.

As it is apparent from Fig. \ref{fig:g-function}, $g(E)$
is a smooth function of its argument. If the shape of the leading order
differential cross section is such that
only a certain range of energies dominates the above integral, $g(E)$ can
be expanded
around some point $E_0$ inside that range. Such an expansion leads to the
following series for the total
cross section:
\beqa
\label{eq:cross-sec-series-full}
\sigma_{CC}&=&\sigma^{\rm Tree}_{CC}\left(1+{\alpha\over\pi} g(E_0)\right)
\nonumber \\
&+&{\alpha\over\pi}\Biggl( g^\prime(E_0) \int_{m_e}^{E^{max}} (E-E_0)
d\sigma^{\rm Tree}_{CC}(E) \nonumber \\
&+&{1\over 2}g^{\prime\prime}(E_0) \int_{m_e}^{E^{max}}
(E-E_0)^2d\sigma^{\rm Tree}_{CC}(E) + \cdots \Biggr)
\eeqa
If the point $E_0$ is chosen in such a way that
\beq
\label{eq:e-average}
E_0={\int_{m_e}^{E^{max}} E d\sigma^{\rm Tree}_{CC}(E) \over \sigma^{\rm
Tree}_{CC}}\equiv \langle E\rangle
\eeq
then the second term in Eq.(\ref{eq:cross-sec-series-full}) vanishes. Here,
$\langle E\rangle$ represents
the average observed energy calculated with the leading order electron
spectrum. The series now has the form
\beqa
\label{eq:cross-sec-series}
\sigma_{CC}&=&\sigma^{\rm Tree}_{CC}\left(1+{\alpha\over\pi} g(\langle
E\rangle)
+{\alpha\over\pi 2!}g^{\prime\prime}(\langle E\rangle) \langle \delta
E^2\rangle + \cdots \right)\nonumber \\
\langle \delta E^2\rangle &=& {\int_{m_e}^{E^{max}}(E-\langle
E\rangle)^2d\sigma^{\rm Tree}_{CC}(E)
\over \sigma^{\rm Tree}_{CC}}
\eeqa
The above formula shows that the total cross section can be represented as
a series in moments
$\langle \delta E^n\rangle$ of the leading order electron spectrum with
coefficients given by
$\alpha/(\pi n!) g^{(n)}(\langle E\rangle)$, with superscript indicating
the $n$th derivative.
Since we expect the final uncertainty in the cross section to be of the
order of 0.1\% (see
Section \ref{sec:manu-nucl}), the series can be truncated at the leading
term if the following condition holds:
\beq
\label{eq:constraint}
{\alpha\over 2\pi}g^{\prime\prime}(\langle E\rangle) \langle \delta
E^2\rangle \simle 0.1\%
\eeq
Below we consider some examples for which such truncation is possible.

\subsection{Neutrino and Antineutrino Deuterium Disintegrations}

Consider the first two reactions in Eq.(\ref{eq:reactions}). The shape of
the leading order electron spectrum
is dominated by the overlap integral of the wavefunctions of the deuteron
in the initial state
and the two nucleon system in the final state. The overlap integral depends
on the relative
momentum of the two nucleons, and it falls rapidly when the momentum becomes
larger than
the inverse of the scattering length for the final state system (see Ref.
\cite{krmv01} for discussion).
Because of this feature the electron spectrum is strongly peaked near the
endpoint (see {\it e.g.} Ref. \cite{Kelly}), and its
width is determined by the corresponding scattering length:
\beq
\delta E_{N_1N_2} \approx {1\over a_{N_1N_2}^2M_p}
\eeq
where $a_{N_1N_2}$ is the scattering length for the final state containing
nucleons $N_1$ and $N_2$.
The direct evaluation gives $\delta E_{pp}\approx 0.7$ MeV, $\delta
E_{nn}\approx 0.2$ MeV.

For $E_\nu\ge$ 4 MeV (as it is the case for the CC
reaction at SNO) one can use Eq.(\ref{eq:g-total}) to show that
\beqa
\label{eq:g-pr-pr-high}
{\alpha\over \pi}g^{\prime\prime}(E_\nu)&\approx&{0.0037\over E_\nu^2}
\eeqa
for $E_\nu$ is in units of MeV. Therefore, for $E_\nu\simge 5~{\rm MeV}$ we
have for both channels:
\beq
\label{eq:error-NuD}
{\alpha\over 2\pi}g^{\prime\prime}(\langle E\rangle) \langle \delta
E^2\rangle\le 0.01\%
\eeq
which is an order of magnitude better then necessary.

To verify the validity of the truncation procedure we performed a series of
calculations
using Eq.(\ref{eq:exact}) for the neutrino energies $E_\nu\ge 4$ MeV for
both $pp$ and $nn$ final states.
The truncated expression for the total cross-section
\beq
\label{eq:cross-sec-total}
\sigma_{CC}^{\nu+d}=\sigma^{\rm Tree}_{CC}\left(1+{\alpha\over\pi} g(\langle E\rangle)\right)
\eeq

was valid
to within the estimated error Eq.(\ref{eq:error-NuD}). For instance, for the $pp$ final state at $E_\nu=4$ MeV
the exact one-loop radiative correction to the total cross section is equal
to 3.28\%, whereas Eq.(\ref{eq:cross-sec-total}) gives 3.27\% with $\langle E\rangle=2.35$ MeV.

\subsection{The $pp$ fusion reaction}

The electron spectrum for this reaction [the third in Eq.
(\ref{eq:reactions})] is essentially determined
by the final state phase space factor:
\beq
{d\sigma^{\rm Tree}_{pp}\over dE_e}\sim \beta(E_e)E_e^2(E_0-E_e)^2
\eeq
where $E_0$ is the maximum electron energy. Such shape is also typical in
nuclear beta decays.
For the $pp$ reaction in the Sun, the kinetic energy of the
initial state protons may be neglected in computing $E_0$. In that case,
$E_0\approx2M_p-M_d=0.93$ MeV. For the spectrum given by the above
equation, the first two moments
are well approximated by
\beqa
\langle E \rangle&\approx& {{E_0+m_e}\over 2} \nonumber \\
\langle \delta E^2 \rangle &\approx& {(E_0-m_e)^2\over 24}
\eeqa
Also, one can show from Eq.(\ref{eq:g-total}) that for $E_e\le$ 1 MeV one
has to a good approximation:
\beq
{\alpha\over \pi}g^{\prime\prime}(E_e)\approx{0.0031\over m_e(E_e-m_e)}
\eeq
with $E_e$ and $m_e$ in units of MeV.
Therefore
\beq
{\alpha\over 2\pi}g^{\prime\prime}(\langle E\rangle) \langle \delta
E^2\rangle\approx
{0.06{(E_0-m_e)\over m_e}}\%
\eeq
In particular, for $pp$ reaction we get the truncation error estimate of
0.05\%, which is more than satisfactory.

The rate of the  $pp$ reaction in the Sun strongly affects the flux of $^8B$
neutrinos $\Phi(^8B)$
recently detected at SNO \cite{SNO,SNO-NC} The relationship between
$\Phi(^8B)$ and the S-factor for the $pp$ reaction is \cite{solar-x-sect}:
\beq
\Phi(^8{\rm B}) \sim S_{pp}^{-2.6}
\eeq
Since the fractional change in the total $pp$ cross section due to the
radiative corrections
translates into the same change in $S_{pp}$ we
obtain:
\beq
{\delta \Phi(^8B)\over \Phi(^8B)}=-2.6 {\delta S_{pp}\over S_{pp}}=-2.6
{\delta \sigma^{pp}\over \sigma^{pp}}
\eeq
The exact evaluation using Eq. (\ref{eq:exact}) gives 3.87\% while the
first term
in Eq.(\ref{eq:cross-sec-series}) gives 3.86\% for the average
electron energy of 0.67 MeV. We see that the truncation error is well
within the
aforementioned 0.05\% estimate.

To compare our calculation with the result adopted in Ref.
\cite{solar-x-sect} we need
to subtract the \lq\lq inner"
part of the correction to isolate the \lq\lq outer" piece:
\beq
g(\langle E\rangle)^{outer}=g(\langle E\rangle)-g^{inner}
\eeq
To emphasize that the inner correction is independent of the electron
energy we do not write the argument
for $g^{inner}$. We follow the convention adopted in Ref. \cite{garcia} and
identify
the outer piece with the contribution
from Sirlin's function $G(E_m, E)$ (see Ref. \cite{Sirlin-g}) integrated
over the electron spectrum.
To isolate the remaining inner piece it is convenient to use Eq. (29) of
Ref. \cite{Towner}.
Dropping $G(E_m,E)$ on the RHS of Eq. (29) we obtain:
\beq
g^{inner}=2\ln\left({M_Z\over M_p}\right)+0.55
\eeq
where the 0.55 on the RHS is obtained in Ref. \cite{Towner}. It includes
non-asymptotic contributions
from the weak axial vector current (denoted by $C$ in Ref. \cite{Towner})
as well as perturbative
QCD corrections.
The value of the inner radiative correction is evaluated to be 2.25\%.
Since it is independent of the
electron energy we can simply subtract it from the total correction to find
the outer piece. The latter
is, therefore, equal to 1.62\%, which is 0.22\% higher than 1.4\% used in
Ref. \cite{solar-x-sect}.
The difference is likely caused by a slightly lower value of $E_0$ in the
$pp$ reaction compared
to the corresponding
quantity in neutron decay: $E_0^{n\rightarrow p+e+\nu}=M_n-M_p=1.3$ MeV
compared to
0.93 MeV for the $pp$. Since $g(E)$ is a
decreasing function of $E$, smaller $E_0$ in the $pp$ reaction means a larger
correction.
Using the actual value of the radiative correction reduces the prediction for $\Phi(^8B)$ by
0.57\% relative to the present value. Although not very significant, this shift
is comparable to the total theoretical uncertainty of 0.5\% assigned to
$S_{pp}$
in Ref. \cite{solar-x-sect}.

\section{Axial current coupling constant}
\label{sec:gA}

An accurate prediction of the cross section or rate of a
process that depends on the nuclear matrix elements of the
weak axial vector current requires an accurate knowledge
of the axial coupling constant $g_A$
of the nucleon. If radiative corrections are included one must use
a renormalization scheme
consistent with the procedure used to extract
the experimental value of this constant. Here
we briefly review the prescription for evaluating radiative
corrections to the hadronic matrix elements of
the axial current when using
the value for $g_A$ recommended by the Particle Data Group \cite{PDG00}.

At leading order in electroweak perturbation theory
(but to all orders in the strong interaction) the nucleon matrix
element of the charge-raising current  has the form:
\beqa
\label{eq:ew-tree}
\langle N_f, p_f|\sum_k {\bar q}_k\gamma_\mu(1-\gamma_5)\tau^+q_k |N_i, p_i
\rangle
&=&{\bar p}(g_V \gamma_\mu+g_M{i\sigma_{\mu\nu}q^\nu\over {2 M_N}} \nonumber \\
&&-{\stackrel{\circ} {g}}_A\gamma_\mu\gamma_5-g_P{q_\mu\gamma_5})n ~,
\eeqa
where the $g_i$'s represent the corresponding coupling constants,
and the sum on the left hand side runs over all quark flavors. The relevant
form-factors
are evaluated at
$q^2\equiv (p_f-p_i)^2=0$, and only
contributions from the first class currents are included.
The above formula leads to the inverse neutron lifetime
\beq
\tau_n^{-1}\sim V_{ud}^2 G_F^2\left(g_V^2+3{\stackrel{\circ}
{g}}_A^2\right)+\cdots~,
\eeq
where $V_{ud}$ is the first element of the CKM quark flavor mixing matrix,
$G_F$ is the Fermi coupling constant determined through the muon lifetime,
and ellipses
represent the small pseudoscalar and weak magnetism corrections.
In the absence of isospin breaking the conserved vector
current hypothesis (CVC) holds, and $g_V\equiv1$.
We will not discuss corrections to this approach in the following (see {\it
e.g.} Ref. \cite{kaiser}).
On the other hand, the value of ${\stackrel{\circ} {g}}_A$ is not protected
from renormalization
by QCD effects.
For the purpose of this discussion we will call $g_A$ appearing in
Eq.(\ref{eq:ew-tree})
the \lq\lq fundamental" axial coupling and accentuate it by a circle.

In practice, one uses the combination of measurements of the neutron lifetime
and its parity-violating decay asymmetry, which depends
on the ratio ${\stackrel{\circ} {g}}_A/g_V$. If CVC is invoked, one can
also determine $V_{ud}$.
However, for the accurate determination of all involved parameters
the above formula for the neutron lifetime is insufficient and the electroweak
radiative corrections must be taken into account.
The radiative corrections to neutron decay
have been evaluated in a number of papers (see, {\it e.g.}, Ref.
\cite{Sirlin,garcia,yokoo}).
The general result of these papers is that the
neutron differential decay rate, averaged over the neutron spin
and integrated over all variables
except the electron energy, has the form
\beq
{d\Gamma_n\over dE_e} \sim G_F^2 V_{ud}^2
\left\{\left
[1+2\Delta_V(E_e)\right]g_V^2+\left[1+2\Delta_A(E_e)\right]3{\stackrel{\circ} {g
}}_A^2\right\}
\eeq
where $E_e$ is the energy of the emitted electron, and $\Delta_V(E_e)$ and
$\Delta_A(E_e)$ are,
respectively, the radiative corrections to the vector
and the axial vector current contributions to the decay. It
is important to note that these corrections are in general not equal.
However, for the level of precision we are interested in their difference
can be regarded as
a constant independent of the electron energy \cite{yokoo}.
Although the value of this constant
can in principle be computed in Standard Model,
the practical calculation contains hadronic structure
uncertainties \cite{krmv01}.
It is, however, possible to absorb such uncertainties in the modified
definition of $g_A$:
\beqa
\label{eq:g-a-mod}
{d\Gamma_n\over dE_e} &\sim& \left[1+2\Delta_V(E_e)\right]G_F^2
V_{ud}^2 \left(g_V^2+3g_A^{2}\right) ~, \nonumber \\
g_A &\equiv& {\stackrel{\circ}
{g}}_A\left[1+\Delta_A(E_e)-\Delta_V(E_e)\right]={\stackrel{\circ}
{g}}_A(1+\delta) ~.
\eeqa
The advantage of the above definition is that the same
radiative correction $\Delta_V(E_e)$ dominates
the corrections to the rate of superallowed nuclear beta decays,
which are pure Fermi (vector current) transitions.
Its uncertainty contributes to the uncertainty in the value of
$V_{ud}$ extracted from such decays.
At present, the most conservative estimate of the
combined experimental and theoretical
uncertainty in $V_{ud}$ is 0.07\% \cite{PDG00},
which sets the upper bound on the uncertainty
in $\Delta_V(E_e)$. With $\Delta_V(E_e)$ constrained in this way,
$g_A$ defined in
the second line of the above equation can be experimentally
determined from the neutron lifetime
and its decay asymmetry. It is precisely the value for $g_A$
defined in the above equation,
not for the fundamental ${\stackrel{\circ} {g}}_A$,
that is quoted in the Particle Data Group \cite{PDG00}. The current
number is $g_A=1.2670\pm0.0030$.

The above considerations --in particular Eq.(\ref{eq:g-a-mod}) --
show that instead of evaluating the
radiative corrections to the hadronic matrix elements of the axial
current one can use the corrections to the corresponding matrix elements
of the vector current in combination
with the modified axial coupling constant $g_A$.
For the reactions in Eq.(\ref{eq:reactions}), there is no need to know the
fundamental coupling ${\stackrel{\circ} {g}}_A$\footnote{For other
processes, however,
one does require ${\stackrel{\circ} {g}}_A$. In neutral current lepton-nucleon
scattering, for example, one must include only radiative corrections to the
neutral
current amplitude. The use of $g_A$ rather than ${\stackrel{\circ} {g}}_A$
would
erroneously introduce the effects of charged current radiative corrections
to the
neutral current amplitude.}.
Such an approach eliminates most of the
hadronic uncertainties mentioned in
Ref. \cite{krmv01}, where the discussion was given in terms of the
fundamental ${\stackrel{\circ} {g}}_A$.

\section{Nuclear Structure-Dependent Contributions}
\label{sec:manu-nucl}

\begin{figure}
\begin{center}
\includegraphics[width=4in]{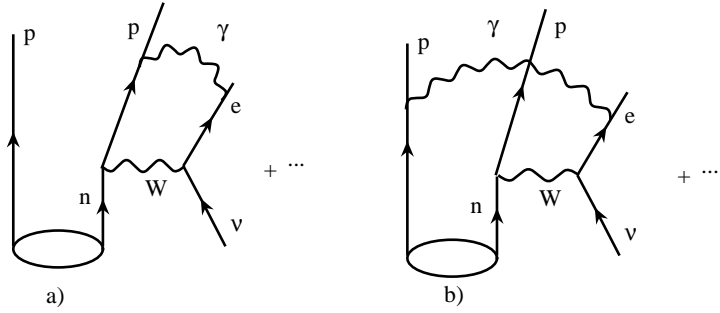}
\caption{One-nucleon (a) and two-nucleon (b) contributions to the radiative
corrections to
CC neutrino-deuterium disintegration. The oval represents the initial state
deuteron.}
\label{fig:rc-types}
\end{center}
\end{figure}

Up to this point, our discussion of radiative corrections has applied to
the one-nucleon approximation;
corrections were computed assuming that both the $W$ boson and the photon
couple to the same nucleon.
This approximation corresponds to graphs in Fig. \ref{fig:rc-types}(a).
There are, however, other
contributions, such as those in Fig. \ref{fig:rc-types}(b). These
contributions depend on the nuclear structure,
which, in general, renders them non-universal.
Below we present arguments suggesting that such corrections should
contribute to the cross
section at the level of 0.1\%.

If the nuclear structure-dependent contributions to the cross section are
not neglected
then Eq.(\ref{eq:g-zero-thresh}) is modified to \cite{towner-c}:
\beq
d\sigma_{CC}(E_{obs})=d\sigma^{\rm Tree}_{CC}(E_{obs})
\left(1+{\alpha\over \pi} \left[g(E_{obs}) + C_{NS}\right]\right) ~,
\eeq
where the quantity $C_{NS}$ represents corrections that depend on nuclear
structure. Its accurate
evaluation requires knowledge of the full nuclear propagator between the
weak and electromagnetic
current insertions. Although possible in principle,
the calculation of this quantity is a complex problem even for such a
simple nucleus as the deuteron.
Before discussing $C_{NS}$ for the deuteron it is instructive to turn to
the case of superallowed nuclear
Fermi $\beta$ decays where the analogous term has already been studied.

Nuclear structure-dependent radiative corrections to superallowed nuclear
$\beta$ decays can be split
into two categories: those induced by weak vector current (VC) and by weak
axial vector current (AC).
For the pure Fermi transitions, the contributions from the VC are suppressed
by powers of the electron energy
or first generation quark mass \cite{abers} and are negligible.
In Refs. \cite{towner-c,towner-new} contributions to $C_{NS}^{\rm F}$
induced by the AC have been evaluated for
Fermi transitions in a number of nuclei using a shell model approach.
$C_{NS}^{\rm F}$ was modeled by contributions analogous to those shown in
Fig. \ref{fig:rc-types}(b).
According to Table II of Ref. \cite{towner-new} the magnitude of
$C_{NS}^{\rm F}$ never exceeds 1.348.

The results presented in Ref. \cite{towner-c,towner-new} are,
of course, model-dependent. However, for pure Fermi transitions
within the Standard Model there exist indications that such approach gives
fairly reliable
values for  $C_{NS}^{\rm F}$. Indeed, these values (along with some other
nucleus-dependent corrections)
successfully bring $ft$-values of various superallowed nuclear beta decays
in agreement
with each other (at about the 0.1\% level) as required by CVC (see
{\it e.g.} Ref. \cite{towner-new} for discussion). This success represents
a non-trivial test of theoretical
nuclear structure corrections in superallowed nuclear beta decays and suggests
that even if the model calculations of such corrections differ from their
actual values the discrepancy must
be a nucleus-independent systematic effect. Although existence of such an
effect might help to solve
the CKM unitarity problem within the Standard Model, the authors of Ref.
\cite{towner-new} argue that
it does not appear plausible under any reasonable circumstances.

These considerations suggest that the existing model calculations for
superallowed
decays produce reliable values for $C_{NS}^{\rm F}$ for a large number of
nuclei with $10\leq A\leq 74$
\cite{towner-new}. Therefore, one might expect that application of similar
techniques to the transitions
in Eq. (\ref{eq:reactions}) will also produce reliable results. Up to terms
suppressed by
electron energy or light quark masses, the dominant nuclear
structure-dependent effects in this case
arise from the VC, rather than the AC as for Fermi transitions\cite{abers}.
Despite this  difference, one might nevertheless argue that an analogous
model calculation of
$C_{NS}^{\rm GT}$ for the
deuteron should be comparable in magnitude to $C_{NS}^{\rm F}$, up to
appropriate scale factors (see
below). According to Ref. \cite{towner-c}
magnitude of $C_{NS}^{\rm F}$ is related to the magnitude of the typical
velocity of a bound nucleon:
\beq
C_{NS}^{\rm F}\sim {\langle p_N \rangle \over M_N}={\langle v_N \rangle
\over c}
\eeq
where $M_N$ is the nucleon mass, $p_N$ is the characteristic nucleon momentum
({\it e.g.} Fermi momentum),
and $v_N$ is the corresponding nucleon velocity . This $v_N$-dependence
follows
directly from the expressions for the nucleon weak and electromagnetic
currents in the non-relativistic
case.

A similar situation occurs for Gamow-Teller transitions. The dominant
nuclear structure-dependent
contribution arises from the antisymmetric correlator of hadronic vector
currents
appearing in one-loop amplitude\cite{abers}:
\beqa
{\cal M}_{NS}^{\rm GT}&\sim& {\bar u}_e \int d^4k{1\over (k^2)^2}
\epsilon^{\alpha\nu\mu\lambda}k_\nu \gamma_\alpha V_{\lambda\mu}
(1-\gamma_5)u_\nu~, \nonumber \\
V_{\lambda\mu}&=&i\int d^4xe^{-ik\cdot x}\langle f|T\{ V_\lambda (x) J^{\rm
EM}_\mu (0)\}| i \rangle~,
\eeqa
where ${\bar u}_e$ and $u_\nu$ are Dirac spinors for the electron and the
neutrino, $V_\lambda$
and $J^{\rm EM}_\mu$ are, respectively the nuclear weak vector current and
the electromagnetic current
operators, and $f$ and $i$ represent the final and the initial nuclear states.
Since two indices
of the $\epsilon^{\alpha\nu\mu\lambda}$ tensor are contracted with the
indices of the two
hadronic vector currents
at least one of these indices must be space-like. Moreover, in the
non-relativistic limit
relevant for low-energy nuclear dynamics, the spatial
components of nucleon vector currents are ${\cal O}(\langle v_N
\rangle/c)$. Hence, we would
also expect $C_{NS}^{\rm GT}$ to scale with $v_N$.

Since the characteristic nucleon momentum in the deuteron is significantly
lower
than that in all nuclei listed in Table II of Ref. \cite{towner-new} our
na{\"\i}ve scaling argument
would give a smaller magnitude for $C_{NS}^{\rm GT}(d)$: $C_{NS}^{\rm
GT}(d)/C_{NS}^{\rm GT}(A)\sim
p_d/p_A$. Here, the quantities describing the deuteron and the nucleus are
accompanied by $d$ and $A$,
respectively. For  atomic number $A\gg 1$ the Fermi momentum is about 300
MeV \cite{preston}.
On the other hand, in the deuteron the corresponding momentum scale is
simply given by its inverse size,
$\gamma$=45 MeV \cite{Kelly}. Taking the largest value for $C_{NS}^F$ from
Table II in Ref. \cite{towner-new} and
rescaling it by the corresponding momentum ratio we
get the following estimate for the maximal expected size of  $C_{NS}^{\rm
GT}$ for the deuteron
disintegration:
\beq
|C_{NS}^{\rm GT}|\sim {45\over 300}\times {\alpha\over \pi}\times
1.348\approx 0.05\%
\eeq
Although such a correction is negligible at present, we cannot rule out
fortuitous few-body structure
effects that might lead to an enhancement. Completion of a detailed
few-body calculation would determine
whether any such enhancement occurs and provide firmer bounds on the
theoretical, nuclear
structure-dependent uncertainty.

\section{Taking the Electron mass to Zero}
\label{sec:mass-singular}

According to KLN theorm \cite{kln}, a properly defined scattering cross section
should have no external mass singularity. In this section we
demonstrate how the result Eq.(\ref{eq:g-total}) is in agreement with this
requirement. We take
$m_e\to 0$ and show that the function $g(E)$ in Eq.(\ref{eq:g-total}) is
finite in this limit.
The results of this section also have a practical application since we
obtain a simple
expression for $g(E)$ valid for $E\gg m_e$, which allows one to evaluate
radiative corrections with accuracy
of about 0.1\% for energies above 1 MeV.

In the limit $m_e\to 0$ we have:
\beqa
\beta(E)&\to& 1-m_e^2/2E^2 \nonumber \\
\ln{1-\beta\over 1+\beta}&\to& 2\ln{2E\over m_e} \nonumber \\
L(\beta)&\to&L\left({1+\beta\over 2}\right)\to L\left({2\beta\over
1+\beta}\right)\to -{\pi^2\over 6} \nonumber \\
L(-\beta)&\to& \pi^2\over 12 \nonumber \\
L\left({1-\beta\over 2}\right)&\to& 0
\eeqa
Using these expression it is straightforward to take the limit $m_e\to 0$
in Eqs. (\ref{eq:gv}) and
(\ref{eq:gb}). We have
\beqa
{\cal A}(\beta)&\to& -\ln^2{E\over m_e} -2\left(\ln 2-{1\over
2}\right)\ln{E\over m_e}
+\ln 2(1-\ln 2)-{\pi^2\over 6}-1 \nonumber \\
g_v(E)&\to& 2\ln\left( {M_Z\over M_p} \right)+\frac{3}{2} ~ {\rm ln}
\left( \frac{ M_p } { m_e } \right)
+\ln^2{E\over m_e}-\ln{E\over m_e} \nonumber \\
&+&\ln 2(1-\ln 2)-{\pi^2\over 6}-{1\over 4}-0.57
\eeqa
The integrals in Eq.(\ref{eq:gb}) can now be evaluated analytically:
\beqa
{1\over 2 E^2 \beta(E) }
\int_{m_e}^E (E-x)\ln\left( {1+\beta(x) \over 1-\beta(x)} \right)dx&\to&
{1\over 2}\left( \ln{E\over m_e}+\ln 2-{3\over 2} \right) \nonumber \\
{2\over E\beta(E)} \int_{m_e}^E { x\beta(x)F(x)-E\beta(E)F(E) \over E-x}dx&\to&
-2\ln{E\over m_e}-2\ln 2 -{\pi^2\over 3}+4
\eeqa
Together with the limit for ${\cal C}(\beta)$
\beq
{\cal C}(\beta) \to -\ln^2{E\over m_e}+\ln{E\over m_e}-\ln 2(1-\ln
2)-{\pi^2\over 6}+1
\eeq
we have
\beqa
g_b(E)&\to& -\ln^2{E\over m_e}-{1\over 2}\ln{E\over m_e} \nonumber \\
&-&\ln 2({5\over 2}-\ln 2)-{\pi^2\over 2}+{17\over 4}
\eeqa
Now we use Eq.(\ref{eq:g-total}) to get
\beq
\label{eq:zero-me}
\lim_{m_e\to 0}g(E)=g_0(E)=2\ln\left( {M_Z\over M_p} \right)
+\frac{3}{2} ~ {\rm ln}  \left( \frac{ M_p } { 2E } \right)
-{2\pi^2\over 3}+3.43
\eeq
This simple formula shows explicitly that there is no mass singularity in
$g(E)$, in
complete agreement with the KLN theorem.

An expression analogous to the RHS of Eq.(\ref{eq:zero-me}) appeared in a
footnote without proof in Ref. \cite{Sirlin}
\footnote{The expression in Ref. \cite{Sirlin} differes from Eq. (\ref{eq:zero-me}) by a factor of
two due to a different convention. We take this difference into account when comparing the two formulae}.
It was given there as an asymptotic formula
for ${\bar G}(E_m)$ in the
limit of large  $\beta$-decay endpoint energy, $E_m$. The function ${\bar
G(E_m)}$, in turn, is the Sirlin's function from Ref. \cite{Sirlin-g}
averaged over the $\beta$
spectrum. The relationship between $g(E)$ and ${\bar G}(E_m)$ can be
constructed from the definition of
the latter:
\beqa
\label{eq:sirling}
{\bar G}(E_m)&=&{\int_0^{E_m} (E_m-E)^2E^2 G(E_m,E) dE\over
\int_0^{E_m}(E_m-E)^2E^2dE}\nonumber \\
&\equiv&2{\int_0^{E_m} (E_m-E)^2E^2 \left( g_0(E)-2\ln\left( {M_Z\over M_p}
\right)-0.55\right)
dE\over \int_0^{E_m}(E_m-E)^2E^2dE}\nonumber \\
&=&3{\rm ln}  \left( \frac{ M_p } { 2E_m } \right)
-{4\pi^2\over 3}+7.1
\eeqa
The last line above coincides with the formula in Ref. \cite{Sirlin} up to
a small constant, which is unimportant
for our discussion here.

In Eq. (\ref{eq:sirling}), the limit $m_e\to 0$ is taken everywhere. The
second line follows from the
first line because the total beta decay rate must be the same in the limit
$E_\gamma^{min}\to\infty$ (the first line)
and $E_\gamma^{min}\to 0$ (the second line). The constant contribution
containing
$\ln M_Z$ represents the \lq\lq inner" radiative corrections, and it must
be subtracted from $g(E)$ if only the
\lq\lq outer" part of the correction is considered.
Since the limit of large $E_m$ must formally coincide with the limit of
vanishing
$m_e$ it is reassuring that both expressions have the same dependence on $E_m$.

Fig. \ref{fig:g-function}
shows that for $E>1$ MeV the functions $g(E)$ and $g_0(E)$ differ by no
more than 0.1\%.
This suggests, for example, that $g_0(E)$ can be used in place of $g(E)$
for the charged current
deuterium disintegration reaction at SNO, where only the events with
$E_{obs}>5$ MeV
were considered \cite{SNO,SNO-NC}. Incidentally, we have for large $E$ ($E$
is in units of MeV):
\beq
{\alpha\over \pi}g_0^{\prime\prime}(E)={3\alpha\over 2\pi E^2}={0.0035\over
E^2}
\eeq
in excellent agreement with Eq.(\ref{eq:g-pr-pr-high}).

\section{Capture Reactions}
\label{sec:conclusions}

Our analysis thus far has emphasized the universal features of radiative
corrections
for low-energy charged current processes in which the charged lepton
appears in the final
state. It is natural to ask whether the corrections relevant to capture
reactions display
similar universality properties. Examples of such reactions include:
\beqa
\label{eq:capture}
e^- + p&\to& n+\nu_e \nonumber \\
p+p+e^-&\to& d +\nu_e \nonumber \\
^7Be+ e^-&\to& ^7Li+\nu_e + (\gamma)
\eeqa
The last two reactions produce $pep$ and $^7$Be solar neutrinos,
respectively. In the last reaction, the
photon emission is due to $^7$Li nucleus decay from an excited state, to
which $^7$Be decays about 10\%
of the time. Here, we show that an expression analogous to Eq.
(\ref{eq:genresult})
applies for such capture reactions, with $E_{obs}\to E_e$ and
$g_{brem}(E_{obs})\to
g_b^{Capt}(E_e, Q)$, where $E_e$ is the initial state charged lepton energy
and $Q$ is the $Q$-value
for the reaction. The presence of $g_b^{Capt}(E_e, Q)$ implies that the
correction factor in this case
is not universal, though the non-universal dependence on $Q$ can be
computed in a straightforward
manner.

As far as radiative corrections are concerned, all reactions of the type
shown in Eq.(\ref{eq:capture})
can be treated in a unified manner. Although at leading order the neutrinos
are monoenergetic
(for a fixed electron energy), bremsstrahlung from the initial state
electron smears the
spectrum to certain extent. Here, we will not study the spectral properties
of the emitted
neutrino but rather focus on the correction to the total emission rate.

It is straightforward to see that the part of the radiative corrections due
to exchange of a virtual photon is the same
as the one contributing to $g(E_{obs})$. Although the bremsstrahlung
contribution is different, it can be easily
obtained from Eq.(\ref{eq:F-func}) with help of crossing symmetry and
time-reversal invariance.
Specifically, one needs to change the sign of
the photon 4-momentum, which is equivalent to changing the sign of
$E_\gamma$ and $x$ in Eq.(\ref{eq:F-func}).
To get the correction to the rate, we write down the analogue of
Eq.(\ref{eq:br-cross})
using the correct phase space. The result is:
\beqa
\label{eq:br-cross-capt}
d\Gamma_{Capt}^\gamma=\Gamma^{\rm Tree}_{Capt}(E_e,Q)\times {\alpha\over \pi}
{(E_e+Q-E_\gamma)^2\over{(E_e+Q)^2}}
F(-E_\gamma,E_e,-x)dxKdK~,
\eeqa
where $\Gamma^{\rm Tree}_{Capt}(E_e,Q)$ is the tree-level capture rate for
a given electron energy and
$Q$-value of the transition.
To study the correction to the total rate we can assume that the
bremsstrahlung is never seen. This leads to the
following expression:
\beqa
\label{eq:capt-brem}
g_b^{Capt}(E_e,Q)&=&\int_{-1}^1dx\int_0^{E_e+Q}KdK{(E_e+Q-E_\gamma)^2\over{(E_e+
Q)^2}}
F(-E_\gamma,E_e,-x) \nonumber \\
&=&2\ln\left( {E_e+Q\over \lambda} \right)
\left[ {1\over 2 \beta(E_e)} \ln \left( {1+\beta(E_e) \over 1-\beta(E_e) }
\right)-1\right]
+{\cal C}(\beta(E_e)) \nonumber \\
&+&{(E_e+Q)^2\over 24 E_e^2}{1\over \beta(E_e)} \ln \left( {1+\beta(E_e)
\over 1-\beta(E_e) } \right)\nonumber \\
&-&{11E_e+2Q\over 3 E_e}\left[ {1\over 2 \beta(E_e)} \ln \left(
{1+\beta(E_e) \over 1-\beta(E_e) }
\right)-1\right]
\eeqa
where the notation is the same as in the discussion of $g(E_{obs})$. It is
apparent that the part that contains
the infrared divergence is the same as before. To get the total correction
to the rate we simply add
the pieces that correspond to the virtual photon exchange (see
Eq.(\ref{eq:outer}) and Eq.(\ref{eq:g-low})):
\beq
g_{Capt}(E_e,Q)=g_v(E_e)^{\kappa\ge\Lambda}+g_v(E_e)^{\kappa\le\Lambda}+g_b^{Capt}(E
_e,Q) ~.
\eeq
It is easy to show that the same formula applies to the positron capture case.

\begin{figure}
\begin{center}
\includegraphics[width=4in]{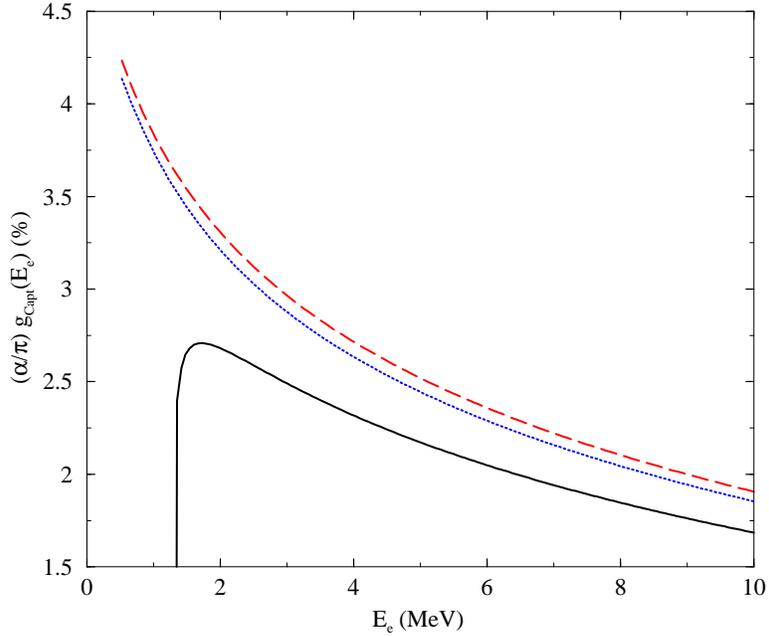}
\caption{The electron energy
dependence of the one-loop radiative corrections to the electron
capture
reactions in Eq. (\ref{eq:capture}):
$p+e^-\to n+\nu_e$ (solid
line), $p+p+e^-\to d+\nu_e$ (dashed line), and
$^7Be+ e^-\to ^7Li+\nu_e$
(dotted
line).}
\label{fig:capture}
\end{center}
\end{figure}

To illustrate the dependence of $g_{Capt}$ on the $Q$-value of the
transition we plot in
Fig. \ref{fig:capture} the value of this function for a range of electron
energies for three cases corresponding
to the transitions in Eq.(\ref{eq:capture}):
\beqa
Q_{p}&=&M_p-M_n=-1.3 {\rm~MeV~,} \nonumber \\
Q_{pep}&=&2M_p-M_d=0.93 {\rm~MeV,~ignoring~pp~kinetic~energy~,} \nonumber \\
Q_{^7Be}&=&M_{^7Be}-M_{^7Li}=0.35
{\rm~MeV,~transition~to~the~ground~state~of ~^7Li~.} \nonumber \\
\eeqa
Note the turnover on the solid curve ($Q_p$=-1.3 MeV). Generally, the
turnover always appears
if $Q+m_e<0$. In such case  the electron must have kinetic energy of at
least $T_e^{min}=-Q-m_e$ for
the capture to occur. The one-photon bremsstrahlung correction formally
diverges at the threshold
$E_e+Q\to 0$, as evident from Eq.(\ref{eq:capt-brem}). This indicates that
one-photon
approximation breaks down, and contributions with arbitrary number of both
real and virtual soft
photons must be resumed to obtain a meaningful answer. In practical
applications
this nicety will rarely be of any importance, however, since the
perturbation series breaks down only if
\beq
\label{eq:breakdown}
{\alpha\over \pi}\ln\left({m_e\over E_e+Q}\right)>1,~{\rm or}~E_e+Q<m_e
e^{-{\pi\over \alpha}}
\eeq
In practice, the spectrum of the initial state electrons is always
many orders of magnitude wider than the last term in Eq.(\ref{eq:breakdown}).
Since the leading order capture rate is suppressed by a factor of $(E_e+Q)^2$
the contribution from the immediate vicinity of the threshold to the
average rate is negligible.

In summary, the radiative corrections for capture reactions are not
described by a universal function.
The initial state bremsstrahlung $\gamma$ -- which is not generally
detected -- cannot radiate away more
of the electron energy than required to make the reaction occur, thereby
introducing a $Q$-dependence
into the radiative corrections. Such considerations do not apply when the
charged lepton appears in the
final state, since in this case the minimum detectable, observed energy is
determined by
the experimental configuration rather the the $Q$-value of the reaction.
The precise value of
energy transferred from the lepton to the hadronic system does affect the
energy-dependence of the
total cross section, but this dependence is the same for both the
leading-order and ${\cal O}(\alpha)$
contributions. Thus, apart from nuclear structure-dependent terms that are
likely negligible for
present purposes, the {\em relative} correction to the tree-level cross
section for the reactions in
Eq. (\ref{eq:reactions}) is described by a universal function. These
features, then, have allowed us to
formulate a unified treatment of radiative corrections for neutrino
reactions that we hope will
help facilitate the analysis of future neutrino property studies.

\begin{acknowledgments}This work was supported in part by the  NSF Grant No. PHY-0071856
and by the U. S. Department of Energy Grants No. DE-FG03-88ER40397 and DE-FG02-00ER4146.
\end{acknowledgments}

\appendix

\section{Closed form for some integrals}
\label{app:integrals}

In Eq. (\ref{eq:gb}) $g_b(E)$ is given in terms of two non-singular integrals over the electron energy. These integrals
can be evaluated in closed form, with the results presented below.  Our evaluation of the integrals is
consistent with the results originally obtained in Refs. \cite{i-fayans,i-vogel}.
In practical calculations,
numerical integration might, in fact, be more efficient than the evaluation of the exact expressions
due to the complexity of the analytic formulas. However, closed form for the integrals may prove valuable if one
is interested in studying the radiative corrections in various limiting cases (such as $m_e\to 0$, {\it etc.}).
For the first integral, we obtain:
\beqa
\int_{m_e}^E (E-x)\ln\left( {1+\beta(x) \over 1-\beta(x)} \right)dx &=& {E^2\over 2} \Biggl[ 
\left({3\over 2}-{\beta^2(E)\over 2} \right)\ln\left( {1+\beta(E) \over 1-\beta(E)} \right)-3\beta(E) \Biggr]
\eeqa
The second integral is more complicated:
\beqa
&&\int_{m_e}^E { x\beta(x)F(x)-E\beta(E)F(E) \over E-x}dx =
-\beta(E)E\left[{1\over 2\beta(E)}\ln\left({1+\beta(E) \over 1-\beta(E)}\right)-1\right] \nonumber \\
&+&{E\over 2} \ln\left({1+\beta(E) \over 1-\beta(E)}\right)
\left[ \ln2\left( 1+{1\over\sqrt{1-\beta^2(E)}} \right)+1 \right] \nonumber \\
&-&{1\over 4}\ln^2\left({1+\beta(E) \over 1-\beta(E)}\right)+L\left({2\beta(E) \over 1+\beta(E)}\right) \nonumber \\
&-&\beta(E) E \left[ \ln2\left(1+{1\over\sqrt{1-\beta^2(E)}}\right) -1\right]
\eeqa
where $F(E)$ is defined in Eq. (\ref{eq:gb}), and $\beta(E)$ and $L(x)$ are defined in Eq. (\ref{eq:functions}).

% One- and Many-body radiative corrections

% Virtual low energy photon exchanges

% Bremsstrahlung graph

% Universal g(E) function

%Various cases of the electron capture

\end{document}